\documentclass[prl,twocolumn]{revtex4}
\usepackage{epsfig}
\usepackage{graphicx}
\usepackage{palatino}
\usepackage{amsmath}
\usepackage{amssymb}
\usepackage{slashed}
\usepackage{epstopdf}
\usepackage{xcolor}
\definecolor{lcolor}{rgb}{0.,0.0,0.}
\definecolor{citcolor}{rgb}{0,0.,0.5}
\usepackage[breaklinks,colorlinks,urlcolor=blue,citecolor=blue,linkcolor=blue]{hyperref}

\newcommand{\beq}{\begin{equation}}
\newcommand{\eeq}{\end{equation}}
\newcommand{\bea}{\begin{eqnarray}}
\newcommand{\eea}{\end{eqnarray}}
\newcommand{\dis}{\displaystyle}
\def\dd{{\rm d}}

\newcommand{\bem}{\begin{multline}}
\newcommand{\eem}{\end{multline}}
\newcommand{\beg}{\begin{gather}}
\newcommand{\eeg}{\end{gather}}

\newcommand{\nn}{\nonumber}

\newcommand{\ben}{\begin{eqnarray*}}
\newcommand{\een}{\end{eqnarray*}}

\newcommand{\eq}[1]{\begin{align}#1\end{align}}
\begin{document}
\title{{\bf Hot spots and the hollowness of proton-proton interactions at high energies \\[0.5cm] }}

\author{
{Javier L. Albacete, Alba Soto-Ontoso }\\[0.2cm]  {\it \small CAFPE and Departamento de F\'isica Te\'orica y del Cosmos,  Universidad de Granada}\\ {\it \small E-18071 Campus de Fuentenueva, Granada, Spain.} \\[0.1cm] {\texttt{ \small albacete@ugr.es, aontoso@ugr.es}}
}

\begin{abstract}
We present a dynamical explanation of the \textit{hollowness} effect observed in proton-proton scattering at $\sqrt s\!=\!7$~TeV. This phenomenon, not observed at lower energies, consists in a depletion of the inelasticity density at zero impact parameter of the collision. Our analysis is based on three main ingredients: we rely gluonic \textit{hot} spots inside the proton as effective degrees of freedom for the description of the scattering process. Next we assume that some non-trivial correlation between the transverse positions of the hot spots inside the proton exists. Finally we build the scattering amplitude from a multiple scattering, Glauber-like series of collisions between hot spots. In our approach, the onset of the \textit{hollowness} effect is naturally explained as due to the diffusion or growth of the hot spots in the transverse plane with increasing collision energy. 

\end{abstract}

\maketitle

\begin{figure}[htb]
\includegraphics[scale=0.56]{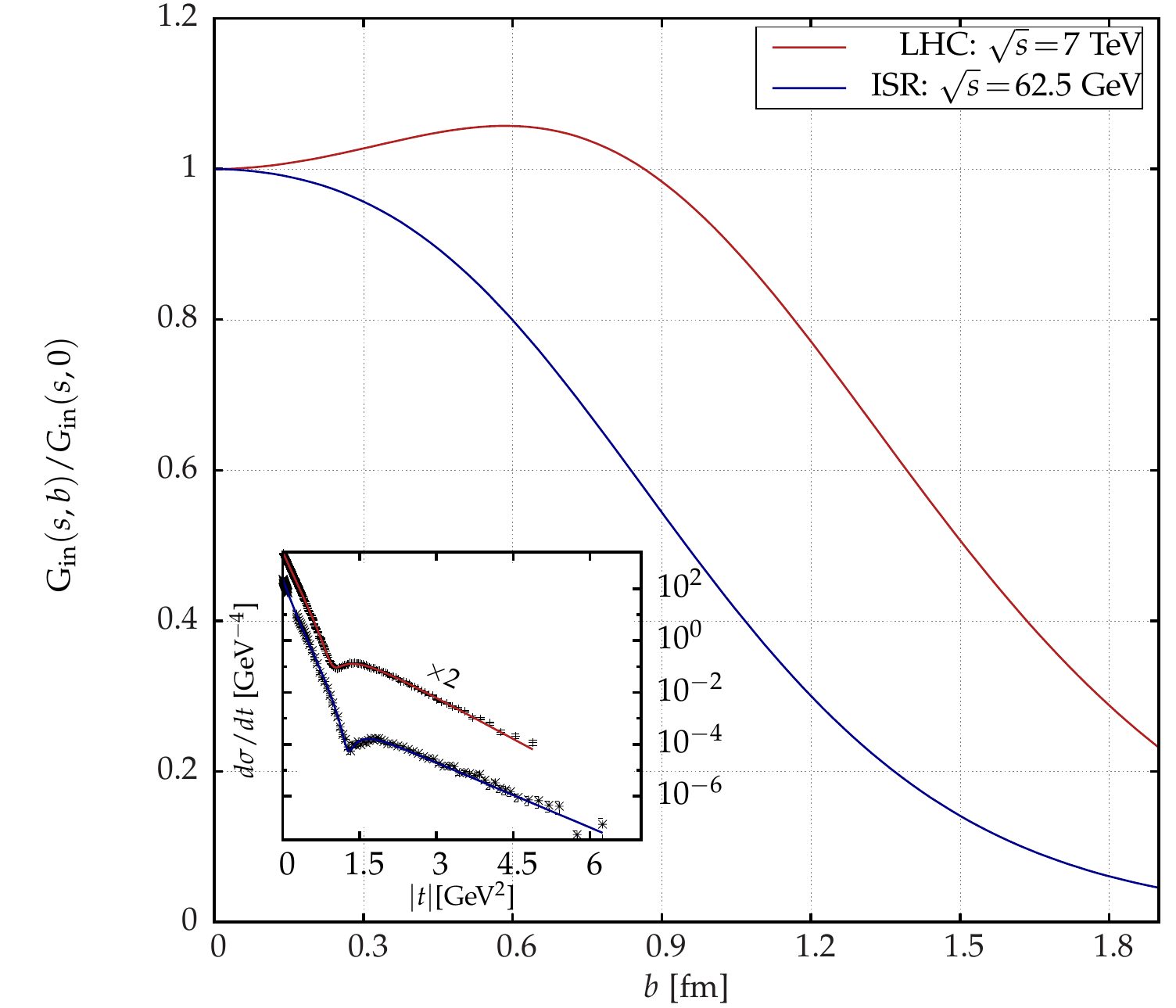}
\vspace*{-0.5cm}
\caption[a]{Normalised inelasticity density, $G_{\rm{in}}$, for LHC and ISR energies as a function of the impact parameter. Sub-pannel: fits to $\dd\sigma_{\rm{el}}/\dd t$ data.}
\label{fits}
\end{figure}

The analysis of experimental data on the elastic proton-proton differential cross-section at collision energy $\sqrt{s}\!=\!7$~TeV measured by the TOTEM Collaboration \cite{Antchev:2011zz} has revealed a new, intriguing feature of hadronic interactions: at high energies, the inelasticity density of the collision does not reach a maximum at zero impact parameter. Rather, peripheral collisions, where the effective geometric overlap of the colliding protons is smaller, are more inelastic or, equivalently, are more effective in the production of secondary particles than central ones. This phenomenon, not observed before at lower collision energies, has been referred to as \textit{hollowness}~\cite{Arriola:2016bxa} or \textit{grayness}~\cite{Alkin:2014rfa,Dremin:2015ujt,Troshin:2016frs} of proton-proton collisions by the authors of the first analyses where it was identified.
Our own independent analysis of LHC and ISR data, to be described below, confirms that the inelasticity density of the collision 
\eq{\label{Gin}
G_{\rm{in}}(s,\vec{b})\equiv\frac{\dd^2\sigma_{\rm{inel}}}{\dd^2b} ={2\rm{Im}}\widetilde{T}_{\rm{el}}(s,\vec{b})-\vert\widetilde{T}_{\rm{el}}(s,\vec{b})\vert^2\,,
 }
where $\widetilde{T}_{\rm{el}}(s,\vec{b})$ is the scattering amplitude in the impact parameter representation, reaches a maximum at  $b\ne 0$ for a collision energy $\sqrt{s}\!=\!7$~TeV, as shown in Fig.~\ref{fits}.
 
The \textit{hollowness} effect challenges the standard geometric interpretations of proton-proton collisions. In particular, it precludes models where the scattering amplitude is built in terms of a positive dependence on the convolution of the density profiles of the two colliding protons. Indeed, it can be shown that the inelasticity density associated to any elastic scattering amplitude thus constructed presents a maximum at zero impact parameter, regardless  how intricate the internal structure of one individual proton may be \cite{Arriola:2016bxa}. These observations suggest that the scattering problem may be best formulated in terms of sub nucleonic degrees of freedom which internal dynamics and correlations should be non-trivial with increasing collision energy. Such is the view adopted in this work, where we consider hot spots to be the effective degrees of freedom in terms of which to discuss the properties of the scattering amplitude.

The idea that the gluon content of the proton is concentrated in domains of small radius $R_{hs}$, much smaller than the proton electromagnetic radius that controls the valence quark distribution $R_{hs}\ll R_p$, is strongly supported by theoretical and phenomenological arguments. Further, lattice QCD calculations  confirm the smallness of the correlation length of the gluon field strengths inside hadrons~\cite{DiGiacomo:1992hhp}. Such domains of high gluonic density have been dubbed \textit{gluonic drops} or \textit{hot spots} in the literature. While the existence of hot spots inside hadrons is 
widely accepted, the  debate on their ultimate dynamical origin remains open. It is commonly assumed that the gluon content of the proton is radiatively generated from valence quarks in DGLAP or BFKL-like cascades. In this view  hot spots relate directly to the Fock space of valence partons, for which they would provide an effective description. However, the question arises of how and why the resulting glue is confined to a region of small radius. While the intrinsic non-perturbative nature of glue drops has been advocated in~\cite{Shuryak:2003rb,Schafer:1996wv,Braun:1992jp}, the possibility that the hot spots dynamics can be fully described in terms of weakly coupled physics --at least in some kinematic window-- has also been entertained in the literature~\cite{Kovner:2002xa}. More phenomenological approaches combine both views of the problem including non-perturbative gluon masses in order to regulate the long-range Coulomb tails characteristic of perturbative emission kernels~\cite{Kopeliovich:1999am, Schlichting:2014ipa}.

In this work we shall not delve into those arguments and simply assume that hot spots are  adequate  degrees of freedom to discuss inclusive proton-proton scattering at high energies. Our main assumption about their dynamical properties is that collisions between hot spots are fully absorptive over distances smaller than their radius, $R_{hs}$. Hence, in our effective description, hot spots appear as small black disks of average radius $R_{hs}$. We also assume implicitly that their ultimate dynamical origin is correlated to the valence partons, since we model the proton as composed of $N_{hs}\!=\!3$ hot spots.

\begin{figure}
\includegraphics[scale=0.56]{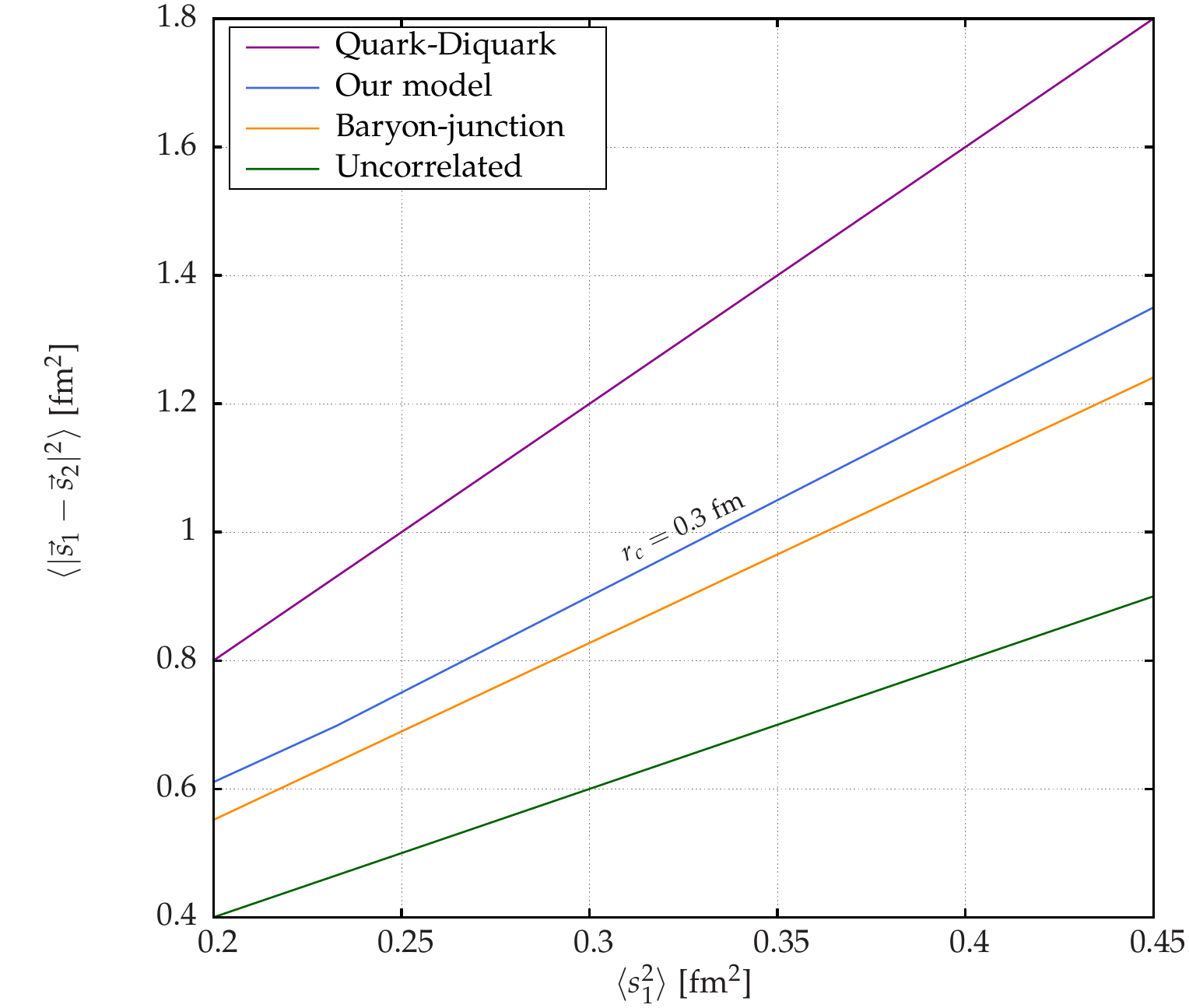}
\vspace*{-0.5cm}
\caption[a]{Mean transverse separation squared between pairs of hot spots as a function of the mean transverse position for different $D(\lbrace\vec{s}_i\rbrace)$.}
\label{s1s2}
\end{figure}

Our goal is to construct the elastic scattering amplitude in proton-proton collisions in impact parameter representation. 
We describe $pp$ interactions as a collision of two systems, each one composed of three hot spots. According to the Glauber model, the natural framework to describe high-energy scattering of composite particles, the elastic amplitude for a collision of particles $A$ and $B$ with the hot spots frozen in transverse positions $\lbrace  \vec{s}_i \rbrace $ has the form:
\eq{\label{expansion}T_{\rm{el}}(\vec{b})=1-\prod_{i=1}^3\prod_{j=1}^3\left[1-\Theta(\vec{b}+\vec{s}_i^A-\vec{s}_j^B)\right]\,, }
where $\Theta$ denotes the scattering amplitude of the $i$-th and $j$-th hot spots interaction and $\vec{b}$ is the impact parameter of the collisions. The physical elastic amplitude is obtained after averaging Eq.~\ref{expansion} over the transverse positions of the hot spots as given by their probability distributions $D(\vec{s}_1,\vec{s}_2,\vec{s}_3)$ in the projectile and target, $A$ and $B$:
\bea
\label{telast}
\widetilde{T}_{\rm{el}}(\vec{b})\!=\!\dis\int\prod_{k,l}\dd^2s_k^{A}\dd^2s_l^BD_A(\{\vec{s}_k^A\})D_B(\{\vec{s}_l^B\})
T_{\rm{el}}(\vec{b})\,.
\eea
The general structure that we shall consider for the joint probability distribution for the transverse positions of hot spots inside a proton has the following form:
\eq{\label{DD}
D(\lbrace\vec{s}_i\rbrace)=C \left(\prod_{i=1}^3 d(\vec{s}_i;R)\right)\times f(\vec{s}_1,\vec{s}_2,\vec{s}_3).
}
The constant $C$ is a normalisation constant to ensure that the probability distribution is normalised to unity: $\int\lbrace\dd^2s_i\rbrace D(\lbrace s_i\rbrace)\!=\!1$. The next term corresponds to the product of three uncorrelated probability distributions for a single hot spot, $d(\vec{s}_i)$. In order to facilitate a full analytical calculation of the scattering amplitude, we shall assume them to be of a gaussian form: 
\eq{\label{d}
d(\vec{s}_i;R)=\exp\left(-\dis s_i^2/R^2\right),
}
where $R$ is the average radius of the $d$ distribution. It should not be confused with the proton radius itself $R_p$. Additionally we also consider the following baseline functional form for the uncorrelated part of the probability distribution:
\eq{\label{dip}
\left(\prod_{i=1}^3 d(\vec{s}_i;R)\right)\rightarrow \int_{0}^{\infty}\dd x~xe^{-x}\dis \left(\prod_{i=1}^3 d(\vec{s}_i;\sqrt{x}R)\right).
}
In the absence of non-trivial correlations the latter form yields, after Fourier transforming back to momentum space, a dipole electromagnetic form factor $\mathcal{F}(t)\!\sim \!1/(1-t^4/R^4)^2$, in better agreement with data than a purely Gaussian one. Nonetheless, the main conclusions of this work are not affected by either choice. Finally, all the correlation structure is encoded in the function $f$ which, by definition, is not factorisable in the spatial coordinates of the hot spots. We write
\eq{\label{corr}
 f(\vec{s}_1,\vec{s}_2,\vec{s}_3)=\delta^{(2)}(\vec{s}_1+\vec{s}_2+\vec{s}_3)\dis\prod_{\substack{{i<j}\\{i,j=1}}}^3\left(1-e^{-\mu\vert\vec{s}_i-\vec{s}_j\vert^2/R^2}\right).}
The $\delta$-function in Eq.~\ref{corr} ensures that the hot spots system is described with respect to the proton centre of mass, thus preventing it from acquiring unphysical transverse momentum. Next, we implement repulsive short-range correlations between all pairs of hot spots controlled by an effective repulsive core $r_c\equiv R/\mu$. In the limit $\mu\to\infty$ we recover the uncorrelated case. While we have no clear dynamical justification for these correlations, their main role in our calculation is to enforce a larger transverse separation between hot spots with respect to the completely uncorrelated case. Indeed, all realistic models for the electromagnetic nucleon form factors entail non-trivial spatial correlations between the constituent quarks: diquark models, where the proton is envisaged as a bound diquark state interacting with the third quark via gluon flux tubes, depict a rod-like structure of a typical string length $l_s\!\sim\!1.5$ fm. In turn, baryon junction models, where the Wilson lines link the three valence quarks at a junction, yield a more triangular structure of the proton. We argue that three-dimensional realisations of both diquark and baryon junction models, when projected onto the reaction plane, produce a similar correlation structure as the two dimensional repulsive core correlations in Eq.~\ref{corr}. In Fig.~\ref{s1s2} we show the average transverse distance between two hot spots yielded by the uncorrelated distribution, corresponding to $\mu\to\infty$, a correlated one with $r_c\!=0.3$ fm and the corresponding value for two three-dimensional triangular distributions projected onto the reaction plane: equilateral and a highly asymmetric isosceles one, which we take as proxies for baryon junction and diquark models. The results for the probability distribution given by Eqs.~\ref{DD}-\ref{corr} provide a good interpolation between the aforementioned, more realistic models of proton substructure. 

In the Glauber formulation of the scattering process adopted here, the main effect of correlations in the transverse positions is to reweight the contribution of the different terms of the multiple scattering series spanned by Eqs.~\ref{expansion} and \ref{telast} with respect to the uncorrelated case. For instance, in baryon junction models, terms where the three hot spots in one proton undergo simultaneous scattering with constituents of the target are strongly suppressed, since the three vertices of an equilateral triangle cannot overlap in the transverse plane, unlike the uncorrelated case. 
The other parameter that controls the amount of effective overlap for different  scattering configurations, and hence their relative contribution to the scattering series, is the hot spot radius. 

To conclude the description of our model we have to specify the elastic scattering amplitude between two hot spots separated a transverse distance $s_{ij}$. Again, in order to facilitate full analytic calculations, we resort to a Gaussian parametrisation: 
\eq{\Theta(s_{ij})=\mathrm{i}\,\exp\left(-\dis s_{ij}^2/2R_{hs}^2\right)(1-\mathrm{i}\rho_{hs})\,.
\label{theta}}
This amplitude can be thought as resulting from the convolution of two gaussian density distributions for a single hot spot, each of radius $R_{hs}$. The absorptive part of Eq.~\ref{theta} is equal to unity for head-on collisions, in line with our assumption of hot spot as small black discs. Although hadronic amplitudes are expected to be mostly imaginary at high energies, we allow for a constant real part $\rho_{hs}$, i.e independent on the momentum transfer, in order to match the non-zero values measured experimentally. It should be noted that all the energy dependence in our model has been left implicit so far. As discussed below, it is encoded in the transverse growth of the hot spot radius $R_{hs}$ with increasing collision energy, that, together with  $R$, $r_c$ and $\rho_{hs}$ are the four  parameters of our model.

Before discussing the results of our model, we briefly present an independent analysis of experimental data on the differential elastic cross section measured at the LHC and ISR at collision energies 7 TeV and 62.5 GeV respectively. 
To describe the data on $\dd\sigma_{\rm{el}}/\dd t=(1/4\pi)\left|T_{\rm{el}}(s,t)\right|^2$ we use the following parametrisation:
\eq{\label{fitT}
{\rm{Im}}T_{\rm {el}}(s,t)&=a_1e^{b_1t}+a_2e^{b_2t}+a_3e^{b_3t}\,,\nn \\
{\rm{Re}}T_{\rm{el}}(s,t)&=c_1e^{d_1t}\,,}
with the fit parameters $a_i$ and $c_i$ subject to the constraints: $\sigma_{\rm{tot}}\!=\!2\dis\Sigma_ia_i$ and $\rho\!=\!\dis\Sigma_i\dis (c_i/a_i)$, where the values of $\sigma_{\rm{tot}}$ and $\rho$ correspond to the experimental measurements given in \cite{Antchev:2011zz,Amaldi:1979kd}. For the LHC case we use the extrapolated  $\rho$ value provided by the COMPETE Collaboration \cite{Cudell:2002xe}, same as the TOTEM collaboration in their data analysis. With this set up we obtain a very good description of $\dd\sigma_{\rm{el}}/\dd t$, $\chi^2/\rm{d.o.f}\sim 1.1\div 2$, and confirm the hollowness effect in LHC experimental data, as shown in Fig.~\ref{fits}. Fits with a larger number of free parameters lead to similarly good data description.

In order to prove that our model actually accounts for the onset of hollowness effect at high energies, we scan the parameter space looking for the presence of a dip of the inelasticity density at zero impact parameter and a monotonically decreasing scattering amplitude, as dictated by data. I.e, we impose the following conditions:


\begin{eqnarray}
\dis\left.\frac{\dd^2 \widetilde{T}_{\rm{el}}(s,b)}{\dd^2 b}\right|_{b=0}<0,\\
\dis\left.\frac{\dd^2 G_{\rm{in}}(s,b)}{\dd^2 b}\right|_{b=0}>0\,.
\end{eqnarray}
We shall refer to the region of parameter space that fulfills the two above conditions as hollowness region.

\begin{figure}
\begin{center}
\includegraphics[scale=0.56]{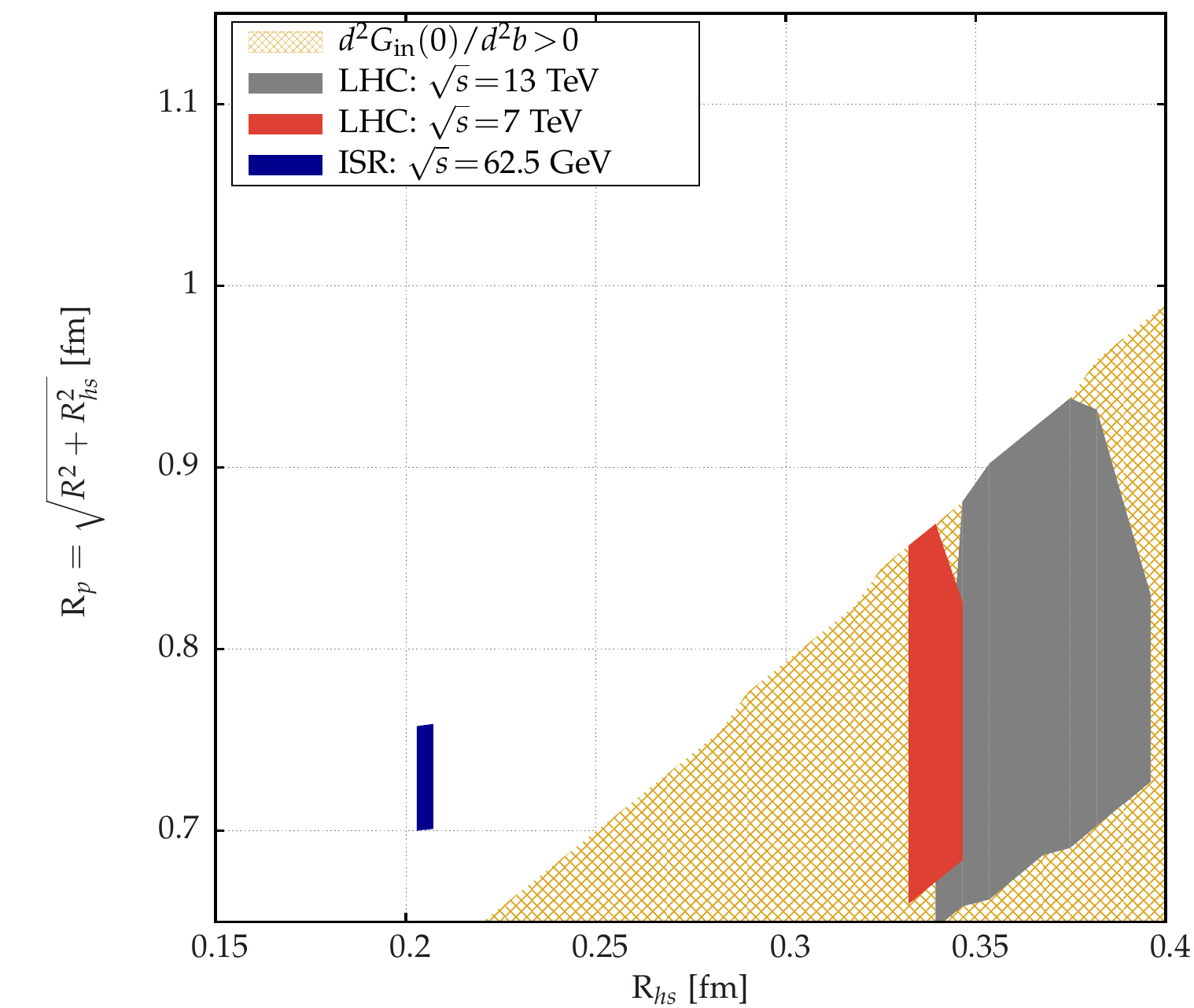}
\includegraphics[scale=0.56]{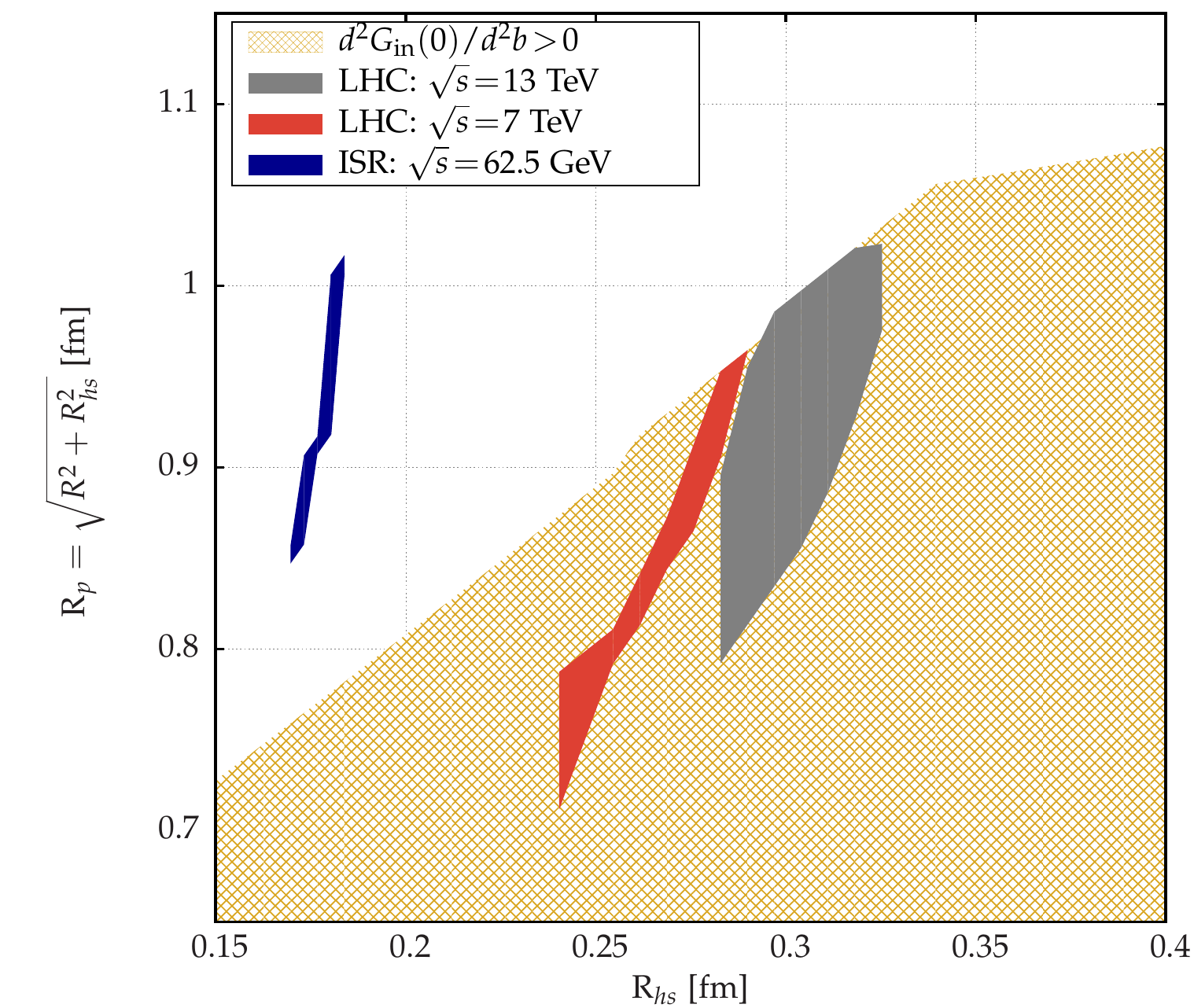}
\end{center}
\vspace*{-0.25cm}
\caption[a]{Hollowness region (filled-dashed) and phenomenologically compatible regions with LHC and ISR data in the ($R_p$,$R_{hs}$)-plane for $r_c\!=\!0.3$ (top) and $0.5$~fm (bottom).}
\label{rc03PS}
\end{figure}

A first, important result is that it is not possible to obtain a growing behaviour of $G_{\rm{in}}(s,b)$ at zero impact parameter in the absence of non-trivial correlations, i.e for $\mu\to \infty$ in Eq.~\ref{corr} or, equivalently for zero correlation distances $r_c\!=\!0$. Indeed, in this case the full calculation of the physical scattering amplitude simplifies and it can be shown that $G''_{\rm{in}}(b=0) <0$ in all cases, regardless the specific functional form for the uncorrelated weights in Eq.~\ref{d}.

In turn, for non-zero values of the correlation length $r_c\!<\!R$, we find a wide region of the parameter space compatible with the hollowness effect. In Fig.~\ref{rc03PS} we show the hollowness region --represented as a dashed area in the plots-- in the $(R_p,R_{hs})$-plane for $r_c\!=\!0.3$ (top) and 0.5 fm (bottom), where we have defined $R_p^2\equiv R^2+R_{hs}^2$, the effective proton radius resulting from the convolution of the hot spots distribution with their own density distribution. Results in both plots were obtained with $0.05\!<\!\rho_{hs}\!<\!0.15$. We observe that the hollowness region enlarges with increasing correlation distance $r_c$. For fixed $R_p$ and $r_c$, the hollowness effect kicks in at some finite value of the hot spot radius $R_{hs}$. For instance, fixing $R_p$ to the value of the measured proton charge radius $R_p\approx 0.88$~fm the hollowness region starts at $R_{hs}\ge 0.34$ and $0.24$~fm for $r_c\!=\!0.3$ and $0.5$~fm respectively. The results shown in Fig.~\ref{rc03PS} correspond to the pure gaussian distribution built from Eq.~\ref{d}. The use of the dipole-like distributions in Eq.~\ref{dip} leads to the same qualitative conclusions on the appearance of the hollowness effect and to very similar quantitative estimate of the corresponding hollowness regions.   

In order to ensure the compatibility of our results with other global features of experimental data, we now explore the phase space region of our model that is compatible with the measured values of the total cross section and the ratio of real and imaginary parts of the scattering amplitude at LHC and ISR energies:
\bea  
\sigma_{\rm{tot}}&=&2{\rm{Im}}T_{\rm{el}}(s,0)=2\dis\int\dd^2b~{\rm{Im}}\widetilde{T}_{\rm{el}}(s,\vec{b})\, \\
\rho&=&\displaystyle\frac{{\rm{Re}}T_{\rm{el}}(s,0)}{{\rm{Im}}T_{\rm{el}}(s,0)}, 
\eea
with $\sigma_{\rm{tot}}=43.32\pm 0.23$~mb, $\rho=0.095\pm 0.018$ at 62.5 GeV (ISR) \cite{Amaldi:1979kd} and $\sigma_{\rm{tot}}=98.3\pm 2.8$~mb, $\rho=0.14^{+0.01}_{-0.08}$ at 7 TeV (LHC).
Upon imposing these further phenomenological restrictions we see how the phase space region phenomenologically compatible with ISR data falls outside the hollowness region. In turn, the subspace compatible with LHC data at 7 TeV fully overlaps with it, both results in perfect agreement with empiric observations. These phenomenologically allowed regions are represented in Fig.~\ref{rc03PS} as dark solid areas. It is also shown the subspace of parameter space compatible with the COMPETE predictions $\sigma_{\rm{tot}}\!=\!111.5\pm10$ mb and $\rho\!=\!0.14^{+0.01}_{-0.08}$ for collision energy 13 TeV which, same as for 7 TeV, is fully contained within the hollowness region. We hence predict that the hollowness effect should also be observed for the collision energy of the Run II at the LHC,13 TeV, provided the COMPETE predictions hold.    

We hence conclude that the main dynamical process underlying the onset of the hollowness effect is the transverse diffusion or growth of the hot spots with increasing collision energy, which is the main result of this work. 
Further, the measured growth of the total proton-proton cross section can be simultaneously accounted for by the same mechanism. Presumably, other soft, genuinely non-perturbative contributions to the cross section may also have influence in the scattering amplitude. However, we have tested that our main conclusions are not affected if we add an additional, energy independent gaussian contribution to Eq.~\ref{telast}, provided that this new soft component contributes less than a 50\% or 25\% of the total cross section at ISR and LHC energies respectively.  

In summary, we propose that the explanation to the rather counterintuitive hollowness effect --whereby proton peripheral collisions are more destructive than central ones at high energies-- lies in the interplay between the different internal scales of the proton: proton radius, hot spot radius and transverse correlation length. The relative enhancement of the destructive interference terms in the multiple scattering series  --known as \textit{shadowing} corrections-- induced by non-trivial probability densities for the hot spots transverse positions and the swelling hot spots radius with increasing energy yield the observed depletion of the inelasticity density in central collisions. These effects may have observable consequences in other sets of experimental data on proton collisions. Arguably, they could impact significantly the interpretation of data specially sensitive to the initial collision geometry, like the correlation and flow analysis of proton-proton collisions and the possible production of small drops of Quark Gluon Plasma in such collisions, a highly debated topic nowadays. 

\section{Acknowledgements} We are grateful to Enrique Arriola for many illuminating discussions. This work is funded by a FP7-PEOPLE-2013-CIG Grant of the European Commission, reference QCDense/631558, and by Ram\'on y Cajal and MINECO projects reference RYC-2011-09010 and FPA2013-47836.

\end{document}